\documentclass{new_tlp}

\usepackage[utf8]{inputenc}
\usepackage[T1]{fontenc}

\usepackage[english]{babel}

\usepackage{graphicx}
\usepackage{url}
\usepackage{fancyvrb}

\usepackage[babel]{csquotes}

\usepackage{amsmath}
\usepackage{amsfonts}
\usepackage{amssymb}

\usepackage{listings}
\AtBeginDocument{}

\usepackage{xcolor}
\usepackage{booktabs}
\usepackage{wrapfig}
\usepackage{microtype}

\usepackage{hyperref}
\usepackage[capitalise,noabbrev]{cleveref}
\usepackage{natbib}

\usepackage{comment}
\usepackage{textcomp}

\newcommand{\ignore}[1]{}
\newcommand{\condtext}[2]{#1}

\newcommand{\prob}{ProB}

\newcommand{\probcli}{\texttt{probcli}}
\newcommand{\probtcl}{\prob{}~Tcl/Tk}

\newcommand{\probtwoui}{\prob{}~2~UI}

\newcommand{\ciao}{Ciao}

\newcommand{\quintus}{Quintus Prolog}
\newcommand{\sicstus}{SICS\-tus Prolog}
\newcommand{\swi}{SWI-Prolog}
\newcommand{\yap}{YAP}

\newcommand{\atomquote}[1]{\textquotesingle{}#1\textquotesingle{}}
\newcommand{\expectsdialect}{\texttt{expects\_\-dialect}}

\newcommand{\library}[1]{\texttt{library\-(#1)}}

\title{Making \prob{} compatible with \swi{}}

\author[D. Geleßus, M. Leuschel]{DAVID GELESSUS, MICHAEL LEUSCHEL\\
	Institut für Informatik, Universität Düsseldorf, Universitätsstr. 1, 40225 Düsseldorf, Germany\\
	\email{\{dagel101,michael.leuschel\}@hhu.de}
}

\submitted{31 January 2022}
\revised{9 May 2022}
\accepted{xx xxxxx xxxx} 

\begin{document}

\maketitle

\begin{abstract}
	Even though the core of the Prolog programming language has been standardized by ISO since 1995,
	it remains difficult to write complex Prolog programs that can run unmodified on multiple Prolog implementations.
	Indeed, implementations sometimes deviate from the ISO standard
	and the standard itself fails to cover many features that are essential in practice.
	
	Most Prolog applications thus have to rely on non-standard features,
	often making them dependent on one particular Prolog implementation
	and incompatible with others.
	We examine one such Prolog application:
	\prob{}, which has been developed for over 20 years in \sicstus{}.
	
	The article describes how we managed to refactor the codebase of \prob{} to also support \swi{},
	with the goal of verifying \prob{}'s results using two independent toolchains.
	This required a multitude of adjustments,
	ranging from extending the SICStus emulation in \swi{}
	on to better modularizing the monolithic \prob{} codebase.
	We also describe notable compatibility issues and other differences that we encountered in the process,
	and how we were able to deal with them with few major code changes.
\end{abstract}

\begin{keywords}
	Prolog, porting, compatibility, emulation, ProB
\end{keywords}

\section{Introduction}
\label{sec:introduction}

For most of its existence,
the Prolog programming language has been continuously developed and extended by different groups,
often in parallel and somewhat independently from one another.
New implementations of Prolog were (and still are) created regularly,
either derived from other existing implementations
or developed completely from scratch.
Each of these implementations supports a different feature set than all others,
often introducing new extensions to the language and libraries
while changing or removing other features.

To establish a common basic definition of Prolog,
the core of the language was formalized in an ISO standard~\citep{iso_prolog_core},
which is followed by the vast majority of Prolog implementations nowadays ---
although some intentionally deviate from the standard in small or large ways.
Furthermore,
as this standard only specifies the \emph{core} of the Prolog language,
it does not cover the libraries and advanced language features of modern full-featured Prolog systems.
Some of these non-standard features have established themselves as \emph{de facto} standards
that are widely and consistently implemented across many modern Prolog systems.
For many other features,
no clear standard has emerged though,
with different systems often supporting the same feature with different interfaces.
Some features are entirely specific to one system and not supported by other systems at all~\citep{prolog_50_years}.

This situation complicates writing Prolog code that can be run on multiple different Prolog systems.
The ISO standard core language alone is insufficient for many programs,
as it does not provide important features like a module system,
constraint programming,
or even simple standard libraries like \library{lists}.
Most non-trivial Prolog code thus requires non-ISO features provided by individual systems.
As Prolog applications are often only tested on the Prolog system for which they were originally developed,
it is easy to accidentally rely on features and behavior not supported by other systems.
Developers may also make the conscious choice to only support a single system,
in order to reduce development effort
or to make full use of a specific system's unique features.

As a case study,
this paper examines the process of refactoring a large non-portable Prolog application
to make it compatible with other Prolog systems.
The application in question,
\prob{},
is an animator, constraint solver, and model checker for formal specifications of safety-critical systems.
It has been in continuous development since the early 2000s,
based entirely on the \sicstus{} system \citep{DBLP:journals/tplp/CarlssonM12}.
As a result,
the \prob{} codebase makes heavy use of SICStus-specific features and libraries,
and compatibility with other Prolog systems was generally not considered in its past development.

Recently,
we have begun refactoring the \prob{} codebase
to make it compatible with more than just \sicstus{},
with a specific focus on supporting \swi{}~\citep{wielemaker:2011:tplp}.
Our goal is to support both Prolog systems in a single codebase ---
we are not planning to drop support for \sicstus{},
but we also do not want to maintain separate branches for the two systems.

\subsection{Motivation}

Our effort to make \prob{} compatible with a Prolog system other than \sicstus{} is motivated by two main factors.

\subsubsection{Double toolchain and reliability}
\label{subsubsec:doublechain}

\prob{} is being used by several companies for safety-critical applications \citep{DBLP:conf/fmics/ButlerKKLLMV20}, e.g., in the development of railway systems
as a tool of class \textbf{T2} as defined by the European norm EN~50128.%
\footnote{
A tool of class T2 ``supports the test or verification of the design or executable code,
where errors in the tool can fail to reveal defects but cannot directly create errors in the executable software''~\citep[Sect. 3.1.43]{EN50128}.
}
However, we strive for \prob{} to be also used as a tool of class \textbf{T3}.%
\footnote{I.e., a tool
that ``generates outputs which can directly or indirectly contribute to the executable code (including data) of the safety
related system''~\citep[Sect. 3.1.44]{EN50128}.}
The use of a non-mainstream language like Prolog is one obstacle for T3 certification of \prob{}, among other aspects.
Although \prob{} features an extensive test suite
that will often detect Prolog system bugs \citep{fide14toolchain},
this risk can be reduced even further
by verifying the results using a second, independent Prolog implementation.
One implementation acts as the primary toolchain, the other as a secondary toolchain validating the output of the primary
toolchain.
Only if the two outputs match can the output of the tool be safely used.

\subsubsection{Free and open-source toolchain and long-term support}

\prob{} is free and open-source software.
However,
because of its dependency on \sicstus{},
which is commercial and closed-source software,
a \sicstus{} license is required to use the \prob{} code.
Users without such a license have so far been limited to using official pre-compiled builds of \prob{}.
By supporting a free Prolog system such as \swi{},
it becomes possible to build and run \prob{} from source without any commercial components,
which makes it easier for outside users to work with the source code.

An open-source Prolog system also allows developers
to debug and fix bugs in the Prolog implementation
without assistance from its upstream developer.
Some applications of \prob{} require a guarantee of long-term support (ranging up to 20 or 30 years),
in which case it is important to not depend on a single upstream developer for support.

\section{Prolog systems and standardization}

The core Prolog language is formalized since 1995 in the \citeauthor{iso_prolog_core} standard
(last revised in 2017),
which specifies Prolog's syntax and semantics,
including control structures and basic built-in predicates.
Practically all Prolog implementations follow this standard to some degree ---
many aim for full compliance,
but others intentionally deviate from the standard in small or large ways
to support old non-standard code or allow implementing new language features.
However, even systems that fall into the latter category
generally try to stay ISO-compatible
unless there is a specific reason for deviating from the standard.

A second part to the standard was published in 2000 and describes a module system for Prolog.
This part of the standard has been largely unsuccessful ---
as of 2022,
aside from one modern implementation\footnote{Amzi!\ Prolog + Logic Server --- \url{https://amzi.com/AmziPrologLogicServer/}},
most systems instead use some variation of the \quintus{} module system
or a different non-standard module system~\citep{10.1007/11799573_6}.
Due to this lack of popularity,
discussion of “the ISO standard” usually only refers to part 1 of the standard
unless part 2 is explicitly mentioned.

Work is ongoing on a technical specification
that formalizes the syntax and semantics for definite clause grammars (DCGs),
which are supported by most modern Prolog systems.
The latest draft version of this TS was published in August 2021~\citep{iso_prolog_dcgs_draft}.

Outside of the ISO standardization process,
a group of Prolog system developers started the Prolog Commons initiative\footnote{\url{https://www.prolog-commons.org/}}
in an effort to establish a common set of standard libraries across Prolog systems.
As of 2022, this initiative appears to be inactive,
with the reference manual being last updated in 2013.
Some portability primitives developed as part of the initiative are now widely supported,
notably conditional compilation and the Prolog flags \texttt{dialect} and \texttt{version\_info} (see \cref{subsec:condcomp}).
On the other hand,
the proposed libraries have not been adopted by any Prolog system as part of its standard library,
although a few systems implement some Prolog Commons predicates in different modules.

No other aspects of Prolog have been formally standardized.
This includes most standard library modules,
as well as advanced language features like term expansion, coroutining, attributed variables, and mutable data.
For some of these features,
certain de facto standards have evolved,
like the widely implemented coroutining predicates \texttt{dif/2}, \texttt{freeze/2}, and \texttt{when/2} (see \cref{subsubsec:coroutining}).
Many other features are not supported as consistently though,
such as attributed variables,
for which there are two similar but incompatible APIs (see \cref{par:attvars}).

In the remainder of this article,
we will specifically focus on \sicstus{} and \swi{},
the two Prolog systems directly relevant to the \prob{} compatibility effort.
Both are mature Prolog implementations
that have been in development since the 1980s
and continue to be actively updated and maintained.
Although the two systems have been developed independently,
both take influence from the Edinburgh/\quintus{} tradition
and offer an overall similar set of language features,
built-in predicates,
and standard libraries.

\sicstus{} offers full conformance to the core ISO Prolog standard.
In contrast,
\swi{} no longer aims to be ISO-compliant,
and in fact introduced certain changes
that are incompatible with the standard and break long-standing tradition.
Specifically,
\swi{}~7 changed the term format of lists
and replaced the traditional meaning of double-quotes with a new string data type.

\sicstus{} is commercial, closed-source software,
developed by Mats Carlsson, Per Mildner, and others at RISE Research Institutes of Sweden.
\swi{} on the other hand is free and open-source,
with development taking place in a public GitHub repository,
primarily by the original author Jan Wielemaker,
although outside contributions are also accepted.

\section{\prob{} architecture}

The core of \prob{} consists of a relatively large Prolog codebase,
made up of about 400 files containing over 150k lines of code (see \cref{tab:prob_source_stats}),
along with a large test suite of almost 7\,000 unit tests and more than 2\,000 integration tests.
Although development began in the 1990s,
\prob{} is still actively developed and targets current versions of \sicstus{} (4.6 and 4.7).
The codebase is fully modularized,
but especially its core modules are tightly interconnected.

In addition to the main Prolog code,
\prob{} includes a number of libraries implemented in C and C++,
which are used to interface with external native libraries
and to implement certain performance-critical code.
\prob{} can also call various external tools,
mostly to support additional specification languages.
Most of these non-Prolog components are optional ---
with the notable exception of the Java-based B parser
which is required for most \prob{} features.

For end users,
\prob{} provides multiple different user interfaces:
a command-line interface (\probcli{}) for batch verification and data validation,
and a set of GUIs (\probtcl{}, \probtwoui{}) for interactive animation, visualisation, and verification.
All of these user interfaces share the same Prolog core,
but interact with it in different ways.
\probcli{} and \probtcl{} are also implemented in Prolog and can call the core directly.
\probtwoui{} is instead implemented in JavaFX and runs as a separate process,
communicating with the core of \prob{} via a network socket.

\begin{table}[t]
	\centering
	\caption[\prob{} source code statistics]{\prob{} source code statistics}\label{tab:prob_source_stats}
	
	\begin{tabular}{l|c|c|c}
		\hline
		& Files & Code lines & Comment lines \\
		\hline
		Core (Prolog) & 165 & 85\,680 & 15\,651 \\
		Extensions (Prolog) & 237 & 59\,358 & 10\,570 \\
		Extensions (C, C++) & 84 & 78\,784 & 1\,691 \\
		GUI (Tcl/Tk) & 23 & 32\,803 & 2\,753 \\
		\hline
	\end{tabular}
	
	{\small (as of \prob{}~1.11.1, released 2021-12-29)}
\end{table}

\section{Timeline of the port}

Before the compatibility effort began,
the \prob{} code was only developed and used on \sicstus{}
and relied on various SICStus-specific language features.
This made even basic compatibility with another Prolog system a challenge ---
for example,
\swi{} could not even parse many of the \prob{} files in their original form,
because of SICStus syntax extensions and minor parser differences.

To evaluate the feasibility of the project,
we began a proof-of-concept port in a separate branch of the \prob{} code.
Expecting that the first porting attempt would be suboptimal and require later reworking,
we did not focus much on maintainability and retaining \sicstus{} compatibility.
Rather,
our main goal with this proof-of-concept port
was to gain an understanding of the basic issues that needed to be solved,
and to decide what compatibility mechanism or library to use.
After about 8 days of work (part-time by a single developer),
we had adjusted the code sufficiently
so that the \prob{} command-line REPL could be started on \swi{}
and simple expressions could be evaluated.

After this initial porting attempt,
we started mostly from scratch
to implement \swi{} compatibility cleanly and remaining compatible with \sicstus{}.
At this point,
we decided to use the \swi{} dialect emulation mechanism (\cref{subsec:expects_dialect}),
which we were originally hesitant to use,
because it is not natively supported by \sicstus{}.
This compatibility issue was easily worked around though,
and the emulation proved to be very helpful
with replacing much of the manual compatibility code
that was needed in the first proof-of-concept port.

With our second port,
it took roughly two months of part-time development
to achieve basic REPL functionality on \swi{}.
After this,
we focused on making all of \prob{}'s basic unit tests pass on \swi{},
which took about four months of part-time work in total.
A significant portion of the development time was spent on developing the \sicstus{}~4 emulation in \swi{} (see \cref{subsec:expects_dialect})
and reporting or fixing SWI-specific bugs and compatibility issues that we encountered.
All of our additions and patches have been submitted upstream,
and most of them are included in the \swi{}~8.4 stable release from September 2021.
We hope that this will reduce the effort needed for other developers looking to port SICStus~4 code to \swi{}.

\section{Porting process}
\label{sec:development}

\subsection{Conditional compilation}
\label{subsec:condcomp}

An important building block for dealing with Prolog system incompatibilities
is the de facto standard \emph{conditional compilation} mechanism
based on the directives \texttt{if(...)}, \texttt{elif(...)}, \texttt{else}, and \texttt{endif}.
It is commonly used to check whether a predicate, library, or other feature exists (see \cref{subsubsec:feature_checks})
and then choose an appropriate code path depending on the features and APIs supported by the running Prolog system.
For example,
the following code defines a predicate to initialize the random number generator,
in a way that is compatible with both the \sicstus{} and \swi{} random number APIs:

\begin{verbatim}
:- if(predicate_property(set_random(_), _)). % SWI-Prolog
  set_new_random_seed :- set_random(seed(random)).
:- else. % SICStus Prolog
  :- use_module(library(random), [setrand/1]).
  set_new_random_seed :- now(TimeStamp), setrand(TimeStamp).
:- endif.
\end{verbatim}

In fact,
\texttt{if(...)} conditions can be arbitrary Prolog goals,
allowing for more complex feature checks as well.

Using the Prolog flags \texttt{dialect} and \texttt{version\_data},
which originated from the Prolog Commons initiative and are now widely supported,
it is also possible to check the name and version of the running Prolog system.
Many conditions can be written either using dialect/version conditions or using feature checks as described above.
Feature checks are almost always preferable,
as they better describe the intended meaning of the condition
and do not need to be manually updated for new releases or other Prolog systems.
Hard-coded version checks sometimes cannot be avoided though
if there are behavior differences that a feature check cannot detect reliably or safely.

\subsection{The \swi{}/\yap{} emulation mechanism}
\label{subsec:expects_dialect}

Conditional compilation can be used to work around nearly all API differences and other compatibility problems.
While this works well in cases where only a few incompatible features need to be used,
it does not scale well for a large non-portable codebase like \prob{},
as every non-portable call in every source file would have to be wrapped in a conditional block.
This can be solved to some extent by extracting the conditional code into a shared module,
but this still requires manual changes to each source file that uses non-portable predicates.

An alternative solution that largely automates this process
is the dialect emulation mechanism supported by \swi{} and \yap{}~\citep{10.1007/978-3-642-18378-2_8}.
The basic concept of this mechanism is that Prolog code declares the “dialect”,
i.e.\ Prolog implementation,
for which it is written ---
for example,
the directive \texttt{:- expects\_dialect(yap).} indicates that the following code was written for \yap{}.
If the declared dialect does not match the running Prolog system,
a set of \emph{emulation} libraries is loaded
to replicate the behavior of the original system.
This usually involves (re-)defining built-in predicates and operators,
providing a different set of standard libraries,
and translating unsupported language features using term/goal expansion.

This emulation mechanism is particularly useful for a codebase like \prob{}
that was not originally written with portability in mind.
Because the emulation mimics the original system's behavior as closely as possible,
the original code often only needs few manual adjustments to run on the new system,
compared to pure conditional compilation.
Emulation cannot fully handle all differences though,
and some complex cases still require manual fixes or conditional code.

\sicstus{} itself does not support the \expectsdialect{} mechanism,
but this did not cause significant compatibility problems.
As \prob{} supports SICStus natively,
emulation is only needed on \swi{},
so the \expectsdialect{} directive can be simply skipped on SICStus.

When we began porting \prob{},
\swi{} only provided emulation for \sicstus{}~3 and not version 4.
Although we were able to reuse some of the existing emulation,
we had to implement many parts from scratch for the SICStus~4 emulation,
because of significant differences in the standard libraries between the two versions.
Internally,
\expectsdialect{} is based on plain Prolog modules,
making it easy for Prolog programmers to extend it
and enabling rapid prototyping of new emulation features.

\subsection{Decoupling optional modules from the core of \prob{}}
\label{subsec:split}

\prob{} includes a large number of features to support a diverse set of use cases.
Beyond the core capabilities of animating and verifying B machines,
\prob{} also for example supports visualizing various aspects of the machine,
or working with other formalisms such as TLA\textsuperscript{+} and Z\@.
Not all of these features are needed at all times or for all use cases though.

Previously,
\prob{} has always unconditionally loaded all of its modules,
even when the code in question was not actually called.
This did not cause problems running on \sicstus{},
as almost all modules are pure Prolog code and thus always load successfully.
For the few modules with problematic dependencies,
\prob{} already supported compile-time flags or automatic detection
to disable the module in question if the needed dependency is missing.

When we began supporting \swi{},
this approach of unconditionally loading all features became a problem.
Most of \prob{}'s code initially did not load or even parse successfully on \swi{},
usually because of SICStus-specific syntax and standard libraries,
or C/C++ dependencies that were not made SWI-compatible yet.
Fixing all of these errors at once would not have been realistic.

To address this,
we began restructuring non-essential \prob{} code into separate \emph{extensions}
that can be selectively enabled or disabled.
This is implemented using term expansion:
\prob{} detects \texttt{use\_module} directives that reference unavailable extensions
and replaces them with stub definitions
that safely display an error when called.

We have also introduced a global flag \texttt{prob\_core\_only},
which disables all optional extensions and leaves only required core modules enabled,
to allow building/running a more minimal version of \prob{}.
Most of our work to support \swi{} was done with this flag set,
so that we could first focus on solving compatibility issues in the core
before gradually tackling the optional extensions.


\subsection{Compatibility issues between \sicstus{} and \swi{}}

We will now present some specific notable differences and compatibility issues between \sicstus{} and \swi{}
that we encountered while porting \prob{},
as well as the solutions or workarounds that we used.
\condtext{}{
Many of these points are also explained in more detail in \citet[section 3.5]{prob_swi}.
}

\subsubsection{Feature checks}
\label{subsubsec:feature_checks}

Even though feature checks are commonly needed in compatibility code (\cref{subsec:condcomp}),
they are often tricky to perform in a way that works correctly on multiple Prolog systems.

The ISO core standard defines \texttt{current\_predicate/1} for checking the existence of a given predicate,
but according to the standard,
it only succeeds for \emph{user-defined} predicates.
Although some implementations like \swi{} also allow it to detect built-in predicates,
others like \sicstus{} strictly follow the standard and fail for built-ins,
making this predicate unsuitable for feature checks.
Instead,
the non-standard but widely supported \texttt{predicate\_property/2} can be used.
Although its intended purpose is to query information about a predicate,
it can also be used to check whether a predicate exists at all,
as it fails without error if asked about a nonexistant predicate.

Existence of libraries (and, more generally, source files) can be checked using the \texttt{exists\_source/1} predicate on \swi{} and \yap{}.
This predicate is not supported by \sicstus{} and many other systems,
but can be emulated using the widely supported \texttt{absolute\_\-file\_\-name/3}\footnote{
	\texttt{absolute\_\-file\_\-name(Source, \_, [access(exist), file\_type(source), file\_errors(fail)])}
}.

Existence of arithmetic functions,
such as \texttt{log(Base,X)},
can be checked using \texttt{current\_\-arithmetic\_\-function/1} on \swi{}.
No other Prolog system,
including SICStus,
provides a similar predicate,
but a check such as \texttt{if(catch(\_ is log(2,4), \_, fail))} can be used instead.

\subsubsection{Term output format}

\swi{} has full Unicode support in Prolog source code,
whereas \sicstus{} is limited to ISO~8859-1 characters outside of quoted literals
and requires non-ASCII whitespace (e.g.\ non-breaking space) to be escaped even inside quotes.
This causes issues when terms written by \swi{} are read by SICStus,
as SWI will not quote or escape some characters as needed by SICStus.

\swi{} \texttt{write\_\-canonical} outputs list syntax to avoid exposing the changed list term format.
This can cause issues with restrictive parsers that do not support list syntax,
such as the one used internally by \probtwoui{}.

Both issues can be worked around by setting appropriate \texttt{write\_\-term} options on \swi{}.

Traditionally and on \sicstus{},
term output predicates like \texttt{print} will never quote atoms and functors.
\swi{} will add quotes as needed so that the term can be correctly read back.
This causes problems when using \texttt{print} to display human-readable text from atoms.
\expectsdialect{} automatically reconfigures \swi{} to the SICStus-compatible behavior.

\subsubsection{Term hashing}

Beyond the basic \texttt{term\_hash/2},
both \sicstus{} and \swi{} support options to customize hashing,
but with largely incompatible APIs.
Overall,
the \sicstus{} \texttt{term\_hash/3} is more flexible than the predicates provided by \swi{}.

In the case of \prob{},
we mainly used conditional compilation to select the best available hashing predicate.
As of writing,
some of \prob{}'s hashing calls are still more optimized for SICStus behavior and options,
leading to hash collisions on \swi{} in some cases.

A further problem is that although each system guarantees stable hashes across releases,
the hashes are not interchangeable between the two systems,
and a hash from one system cannot be reproduced on the other.
This is problematic as \prob{} can save hashes for later verification,
expecting them to be reproducible.
We have not found a good solution for this yet ---
we may need to use a non-built-in, but portable hash function implemented in Prolog or C
that produces consistent results on both Prolog systems.

\subsubsection{Coroutining}
\label{subsubsec:coroutining}

Both systems support the common \texttt{when/2}, \texttt{dif/2}, \texttt{freeze/2}, and \texttt{frozen/2} predicates.
Additionally,
\sicstus{} supports \texttt{block} declarations for coroutines,
which are not natively supported by \swi{}.
However,
\expectsdialect{} provides an emulation of \texttt{block},
which we enhanced to meet the needs of \prob{},
specifically to allow querying blocked goals using \texttt{frozen/2}.

\swi{}'s \texttt{frozen/2} was recently made fully compatible with \sicstus{},
but its implementation at first had serious bugs ---
calling \texttt{frozen} on a variable could e.g.\ could cause internal errors in CLP(FD)
or silently make \texttt{when/2} coroutines not execute\footnote{
	See \url{https://github.com/SWI-Prolog/swipl-devel/issues/828}.
}.
These bugs were detected by the \prob{} test suite
and have since been fixed upstream.

\subsubsection{\sicstus{} implementation details}

\prob{} relies on some undocumented SICStus APIs and implementation details,
which naturally causes issues on other systems.
This was often easily solved by switching to an equivalent documented API,
either always or only conditionally when not on SICStus.

In an extreme case,
\prob{} heavily uses implementation details of SICStus \library{avl} ---
it often directly manipulates the undocumented term structure of AVL trees,
for performance gains compared to the documented API\@.
Replacing all dependencies on this internal term structure
would have been difficult and error-prone.
Instead,
we decided to build a custom \library{avl} implementation
that uses the same internal term format as the SICStus version.
Our custom implementation is written in pure Prolog,
based on \library{assoc} of \swi{}.
Although the custom implementation is only needed on \swi{},
it is also fully compatible with SICStus,
where it has similar performance as the built-in \library{avl}.

\subsubsection{CLP(FD)}

\sicstus{} \library{clpfd} provides an extensive API with many features not found in other implementations.
\swi{} \library{clpfd} also has a number of advanced features,
but not as many as \sicstus{}.
In particular,
\prob{} frequently uses the SICStus “FD set” API,
which was not supported on \swi{},
so we contributed a compatible implementation of the API to \swi{}.

There is a notable technical difference between the two \library{clpfd} implementations:
the \swi{} version is implemented in pure Prolog and supports arbitrarily large integers.
In contrast,
the SICStus version is implemented in native code,
resulting in better performance
at the cost of being limited to the tagged integer range.
When this range is exceeded,
SICStus \library{clpfd} throws an overflow error ---
sometimes in hard to predict places,
due to coroutining,
making it difficult to catch overflow errors reliably.
The \swi{} implementation does not suffer from this problem,
but for some use cases it is noticeably slower than SICStus.

\subsubsection{Command-line syntax}

SICStus and \swi{} use different command-line option names and sometimes require a different argument order.
To run \prob{} easily on both systems,
we wrote a wrapper shell script
that supports the most important options
and translates or reorders them as needed for each Prolog system.

\sicstus{} allows passing \emph{system properties} at the command-line
using the syntax \texttt{-Dname=value}.
\prob{} uses this feature to implement compile-time flags,
e.g.\ to enable debugging/profiling features or disable optional extensions.
\swi{} has no equivalent feature,
so we are currently using regular environment variables as an alternative.

\subsubsection{Minor incompatibilities}

Finally,
we want to mention a few incompatibilities that are important,
but not complex,
and were easily solved or worked around.

\paragraph{Operator declarations}

Global on \sicstus{},
but on \swi{} they are module-local and have to be imported/exported or declared globally.
\expectsdialect{} solves this automatically.
A few cases in \prob{} needed manual adjustment,
but those were already questionable ---
e.g.\ using an operator from a module that was never imported.

\paragraph{\swi{} list term format}

Required few changes overall,
as almost all code uses list syntax
and does not rely on the \texttt{\atomquote{.}} list term functor.
The difference was only noticeable in code that inspects free-form terms,
such as term printing and type checking utilities.

\paragraph{Double-quoted literals}

Represents a list of character codes traditionally,
including on \sicstus{},
but is a “true” string on \swi{} by default.
The meaning can be changed using the ISO standard Prolog flag \texttt{double\_quotes} ---
setting it to \texttt{codes} forces the traditional behavior regardless of the default of the Prolog system.

\paragraph{Mutable terms}

\sicstus{} has a specialized mutable term data type.
\swi{} allows mutating any compound term's arguments.
\expectsdialect{} can easily emulate the SICStus mutable term API\@.

\paragraph{Attributed variables}
\label{par:attvars}

\sicstus{} and \swi{} implement two similar,
but incompatible attributed variable interfaces.
\prob{} handles these differences using conditional compilation.
Although the \swi{} attvar interface is less powerful than the \sicstus{} one,
this has so far not impacted \prob{}'s use of attvars.

\paragraph{Simple naming differences}
\label{par:copy_term}

In many cases (too many to list),
\sicstus{} and \swi{} provide identical predicates under different names,
or with the same name in a different library module.
In rare cases,
both systems have identically-named predicates with different behavior,
such as \texttt{copy\_term/2} (does not copy attributes on SICStus)
or \texttt{lists:prefix/2} (arguments are reversed on SICStus).
All of these differences are handled automatically by \expectsdialect{}.


\section{Empirical Results}
\label{sec:experiments}

\subsection{Current state of compatibility}

At the time of writing,
\prob{}'s command-line interface is usable on \swi{},
supporting interactive REPL usage, batch operations,
and running \prob{}'s test suite.
However,
because most non-core parts of \prob{} are currently disabled on \swi{}
(as explained in \cref{subsec:split}),
a number of \prob{} features are not available yet,
such as support for formalisms other than B and Event-B\@.

To estimate the completeness and correctness of \prob{} running on \swi{},
we used \prob{}'s integration test suite,
which contains a total of 2129 tests.
Of these tests,
954 require features that are disabled in core-only mode,
and 47 further tests are currently disabled for other reasons.

This leaves 1128 tests that can currently be executed on both \sicstus{} and \swi{}.
Most of these tests run successfully on \swi{},
but 72 tests (6.4~\%) still fail.
Many of these test failures are because \prob{} on \swi{} is unable to find a solution that it can find on \sicstus{}.
This makes \prob{} on \swi{} less powerful than on \sicstus{},
but does not affect correctness.

However, some test failures on \swi{} are caused by \prob{} reporting false positives,
e.g.\ invariant violations even though the invariant is true.
We were also able to construct a case that produces a false \emph{negative},
where \prob{} on \swi{} does \emph{not} find a known error in the specification
that is detected when running on SICStus.
We are still investigating the underlying causes of these incorrect results.
As \prob{} has previously encountered implementation bugs in \swi{} related to coroutines\footnote{
	Such as \url{https://github.com/SWI-Prolog/issues/issues/105} (which has since been fixed).
},
we suspect that a similar \swi{} bug might be causing these failures,
but they may also be due to remaining compatibility issues in the \prob{} code.

Until these correctness problems are fixed,
\prob{} on \swi{} on its own cannot be relied on yet.
This is mitigated in a double toolchain setup,
where a mismatch with SICStus would be detected
and prompt further investigation of the results.

\subsection{Performance comparison}

Although our \swi{} support is far from finished,
it is complete enough that we can successfully run some complex test cases.
This allows us to compare the performance of a real-world application running on \sicstus{} and \swi{}.
However, the measurements have to be taken with a grain of salt:
\prob{} has been developed and optimized for \sicstus{} for over almost 20 years.
As such, the comparison is somewhat unfair to \swi{}.
We expect that performance will improve in the future
as we continue to test and optimize \prob{} on \swi{}.

To evaluate the performance for individual applications of \prob{},
we selected a few of \prob{}'s tests from the \texttt{codespeed} test category.
This test category is used to detect performance regressions in \prob{}.
The tests in \cref{table:experiments} are further refined into five different categories:
 micro (micro benchmarks),
 cs (constraint solving),
 cbc (con\-straint-based symbolic verification checks),
 mc (model checking), 
 dv (data validation on large data).

The experiments were run from source, using \prob{}'s test runner.
Each experiment is a test in \prob{}'s test database (file \texttt{src/testcases.pl} in the \prob{} source code\footnote{
	The \prob{} source code and example machine files used in the tests are available at:\\
	\url{https://www3.hhu.de/stups/downloads/prob/source/}
}).
The test number can be found in \cref{table:experiments}.
The tests were run using \sicstus{} 4.7.0, \swi{} 8.5.6, and
 \prob{} 1.12.0-nightly ({\small \texttt{8ad8e874}}) 
 on macOS 12.1 on a MacBook Pro (13'' 2019, 2.8 GHz Quad-Core Intel Core i7).
\Cref{table:experiments} contains both walltime and runtime (time excluding garbage collection
or time spent in non-Prolog code).

\ignore{
Experimental results for constraint solving:
\begin{itemize}
\item 49: flat200-90.mch: Load and solve the B encoding of a SATLIB problem
 \item 55: SortByPermutation: Permutation sort in B setting up sorted constraints
 \item 56: GraphColouring: a B model for graph colouring
 \item 221: CrewAllocation: a B model for a crew allocation scheduling problem
 \item SendMoreMoney, Queens, ...  with CLP(FD) and possibly without CLP(FD)
 \item set comprehension computation, ...
\end{itemize}
}

\begin{table}
\caption{Performance benchmark tests of \prob{} \label{table:experiments}}
\begin{footnotesize}
\begin{tabular}{r|rrr|rrr|cl}
\hline
 Test & \multicolumn{3}{c|}{Walltime (ms)} &  \multicolumn{3}{c|}{Runtime (ms)}  & Cat. & Description\\
  ID & SICS & SWI & Factor & SICS & SWI & Factor &   &  \\
 \hline
65 & 46 & 36 & 0.78 & 44 & 27 & 0.61 & micro & Sequence solving\\
23 & 559 & 1015 & 1.82 & 416 & 571 & 1.37 & micro & Sequence concatenation\\
800 & 3366 & 6110 & 1.82 & 1364 & 4446 & 3.26 & cbc & Bosch cruise controller\\
1748 & 29443 & 75056 & 2.55 & 28069 & 69060 & 2.46 & mc & MCCountToMAXINT50000\\
56 & 67 & 202 & 3.01 & 67 & 190 & 2.84 & cs & Graph colouring\\
1635 & 860 & 3212 & 3.73 & 736 & 2949 & 4.01 & dv & ClearSy/Alstom DTVT\\ 
1963 & 1374 & 5203 & 3.79 & 1328 & 4968 & 3.74 & mc & Volvo cruise controller\\
778 & 6500 & 24960 & 3.84 & 5523 & 22992 & 4.16 & dv & Siemens data validation\\
1336 & 2187 & 8947 & 4.09 & 2097 & 8505 & 4.06 & cs & SlotSolver\\
1195 & 82 & 372 & 4.54 & 61 & 355 & 5.82 & cs & Graph isomorphism large\\
255 & 392 & 1905 & 4.86 & 369 & 1679 & 4.55 & mc & Bepi Colombo mode protocol\\
383 & 713 & 4185 & 5.87 & 694 & 4097 & 5.90 & cs & Sudoku hex puzzle\\
414 & 56 & 329 & 5.88 & 43 & 285 & 6.63 & cbc & BPEL2B PurchaseOrder\\
2015 & 4355 & 27568 & 6.33 & 3425 & 26606 & 7.77 & dv & ClearSy Caval\\ 
55 & 238 & 1910 & 8.03 & 226 & 1868 & 8.27 & cs & Sort by permutation\\
378 & 1691 & 14160 & 8.37 & 1609 & 13634 & 8.47 & cs & NQueens as events\\
221 & 232 & 2000 & 8.62 & 228 & 1948 & 8.54 & cs & Crew allocation puzzle\\
1920 & 643 & 6464 & 10.05 & 561 & 6224 & 11.09 & mc & Performance of while loop\\
1745 & 15 & 180 & 12.00 & 15 & 165 & 11.00 & cs & EulerWay\\
1394 & 833 & 11996 & 14.40 & 824 & 10952 & 13.29 & dv & Alstom IXL Lausanne\\
49 & 2396 & 37307 & 15.57 & 2249 & 32460 & 14.43 & cs & SAT test (flat200-90)\\
387 & 1467 & 25982 & 17.71 & 1443 & 24976 & 17.31 & cs & TwoQueensSevenKnights\\
1715 & 627 & 15790 & 25.18 & 620 & 14809 & 23.89 & dv & Alstom while loop test\\ 
1746 & 14 & 581 & 41.50 & 13 & 528 & 40.62 & cbc & JavaCard model\\
40 & 808 & 150964 & 186.84 & 608 & 148040 & 243.49 & micro & Direct product operator\\
 \hline
\end{tabular}
\end{footnotesize}
\end{table}

As one can see in the above table there is a marked difference in performance in favour of \sicstus{}.
For two thirds of the tests, \prob{} on \swi{} is $3\times$--$15\times$ slower,
and there is one outlier with even a $186\times$ slowdown.
On the positive side, the memory consumption (not in \cref{table:experiments}) remains similar.
Furthermore, for applications where a double toolchain (i.e., the dv category; see \cref{subsubsec:doublechain}) is required,
 runtimes are reasonable for a secondary validation toolchain:
 3.84 slower for Siemens data validation or 6.33 slower for ClearSy data validation example.




As to the reasons for the current performance difference, we can only provide a few guesses.
One reason is certainly that \prob{} makes heavy use of \texttt{block} coroutines in its constraint solver;
these directives have efficient built-in support in \sicstus{},
but are translated to regular Prolog code in \swi{}.
One might also think of the JIT in \sicstus{}, but the first JIT version of \sicstus{} actually made \prob{} run slower for some benchmarks and did not result in big gains.
Performance of \swi{} can vary depending on the C compiler and use of profile-guided optimization ---
we have not yet experimented which configuration yields best performance for \prob{}.


\section{Related work}
\label{sec:related}

\citet{10.1007/978-3-642-18378-2_8},
which introduced the \expectsdialect{} mechanism,
also presents a case study in which the Alpino parser suite~\citep{alpino}
was ported from \sicstus{}~3 to \swi{}~6.
The current version of Alpino\footnote{\url{https://github.com/rug-compling/Alpino}}
remains compatible with both Prolog systems,
although it has not been updated to support recent versions of either system
(\sicstus{}~4 and \swi{}~7/8).

Many of the issues described in the Alpino case study
are similar to those that we encountered with \prob{},
such as library differences,
lack of \texttt{block} coroutines in \swi{},
and incompatible GUI libraries.
When porting \prob{},
we were able to build on many parts of \swi{}'s \sicstus{} emulation
that were originally developed for Alpino.
However,
because of version differences,
many parts of the emulation had to be adjusted or rewritten to support \sicstus{}~4.

Logtalk~\citep{DBLP:conf/padl/Moura13},
an object-oriented Prolog derivative,
is itself implemented in Prolog and highly portable,
supporting 15 different “backend” Prolog systems.
Programs written in Logtalk automatically benefit from this portability,
as most Logtalk features are independent of the backend system.
Not all libraries are abstracted this way though ---
e.g.\ CLP(FD) is exposed largely unmodified from the backend.
Although Logtalk provides much broader portability than other solutions like \expectsdialect{},
it is less suitable for porting large amounts of previously non-portable Prolog code,
as Logtalk is a significantly different language from regular Prolog.

Within the ASAP EU project,
the \sicstus{}-based tools \textsc{ecce} and \textsc{logen} \citep{LeuschelEtAl:PEPM06}
were made compatible with Ciao.
At the time,
conditional compilation was not available yet, 
so manual compatibility modules for each system were used instead.

\section{Conclusion and future work}
\label{sec:conclusion}

We presented the process of making \prob{},
which was originally developed only for \sicstus{},
also compatible with \swi{}.
Our motivations were, amongst others, a double toolchain for certification
 and ensuring long term support.
 
Through the use of the \expectsdialect{} mechanism supported by \swi{},
we were able to do so \emph{without} forking the codebase while keeping support for \sicstus{};
 many differences between the two systems were bridged automatically without large manual code changes.
\swi{} initially did not support \sicstus{}~4 emulation,
 but we were able to contribute this support without much difficulty.
Our additions to the emulation have been released as part of \swi{}~8.4
and can be used by other developers looking to port software from SICStus~4.

Yet \expectsdialect{} is not a one-line solution.
There remain many differences between Prolog systems
that cannot be handled automatically
or are better handled by manual refactoring,
and optimizations made for one Prolog system do not always translate well to other systems.
Furthermore,
porting a complex application to a new Prolog system
is likely to uncover bugs in the targeted system ---
even with a popular and actively-maintained one like \swi{}.

We are planning to further improve \swi{} compatibility and performance in \prob{},
with the eventual goal of fully supporting it as a second toolchain in addition to \sicstus{}.
In the future,
we may also investigate supporting other Prolog systems,
such as \yap{}~\citep{costa2012yap} or \ciao{}~\citep{DBLP:journals/tplp/HermenegildoBCLMMP12}.

\paragraph{Acknowledgements}

We would like to thank Jan Wielemaker for his exceptionally quick responses on the \swi{} forum and bug tracker
and for being so open to improving \sicstus{} compatibility in \swi{}.
We also thank Mats Carlsson and Per Mildner for the excellent support provided for \sicstus{} over many years.
Finally, we are grateful to ClearSy, Alstom and Thales for pushing and helping \prob{} to move towards T2 and T3 certification.

\paragraph{Competing interests}

The authors declare none.

\bibliographystyle{acmtrans}
\bibliography{references}

\end{document}